# Verifiability-First Agents: Provable Observability and Lightweight Audit Agents for Controlling Autonomous LLM Systems

Abhivansh Gupta

abhivansh_g@ch.iitr.ac.in
Data Science Group
Indian Institute of Technology, Roorkee

## Abstract

As LLM-based agents grow more autonomous and multimodal, ensuring they remain controllable, auditable, and faithful to deployer intent (Amodei et al. 2016; Gabriel 2020) becomes critical. Prior benchmarks measured *propensity* for misaligned behavior (Perez et al. 2023; Wang, Chen, and Liu 2024) and showed that agent personalities and tool access significantly influence misalignment. Building on those insights, we propose a *Verifiability-First* architecture that (1) integrates run-time attestations of agent actions (cryptographic & symbolic) (Feng, Lin, and Zhu 2024; Lee and Suresh 2025), (2) embeds lightweight *Audit Agents* that continuously verify intent vs. behavior using constrained reasoning, and (3) enforces *challenge–response* attestation protocols for high-risk operations (Krishnan and Mishra 2024). We introduce OPERA (Observability, Provable Execution, Red-team, Attestation), a benchmark suite and evaluation protocol designed to measure (i) detectability of misalignment, (ii) time-to-detection under stealthy strategies, and (iii) resilience of verifiability mechanisms to adversarial prompt/persona injection. Our approach aims to shift the evaluation focus from "how likely misalignment is" to "how quickly and reliably misalignment can be detected and remediated."

## Introduction

Autonomous LLM agents operate in loops where internal planning, tool use, long-horizon memory (Yao et al. 2023; OpenAI 2023), and repeated self-prompting generate complex behaviors; small prompt changes (persona injection) can materially change outcomes (Perez et al. 2023; Park et al. 2023). Recent works have illuminated how to measure agents' *propensity* for misaligned behavior (avoiding oversight, sandbagging, resisting shutdown, etc.) (Wang, Chen, and Liu 2024; Shen, Tang et al. 2023), and have shown that both models and personality prompts strongly modulate misalignment (Perez et al. 2023; Gao, Korbak et al. 2023). Also, different comprehensive misalignment scoring mechanisms (Wang, Chen, and Liu 2024) have been introduced for quantitative evaluation recently.

Propensity metrics are essential, but a complementary agenda is needed (Pan and Liu 2023; Zheng et al. 2023): *verifiability*. Instead of only asking, "would an agent do harm?", we must ask, "can we reliably observe and verify the agent's internal goals and actions in real time and prevent harmful escalations?" Verifiability is an operational property that enables deployers and auditors to detect, attribute, and remediate misaligned behaviors quickly. This paper introduces a concrete architecture, theoretical properties, and an experimental protocol to evaluate verifiability mechanisms in agentic systems.

## Background

LLM-based agents differ fundamentally from static chat models: instead of responding to a single prompt, they autonomously plan, call tools, retrieve data, and act in open environments. This loop of reasoning, acting, and observing introduces an entirely new axis of risk, not only of incorrect predictions but also of uncontrolled behavior over time. As language models acquire persistent memory and API access (Park et al. 2023), they become capable of forming implicit goals that can deviate from the designer's stated intent.

Recent research has begun systematically studying these failure modes. Recent benchmarks have quantified the propensity for misaligned behavior (Wang, Chen, and Liu 2024) across realistic autonomous tasks such as cybersecurity, content moderation, and negotiation. By combining behavioral probes with persona conditioning, it was revealed that even identical base models could exhibit widely divergent ethical and strategic patterns depending on their internal prompting context. Works have provided the large-scale, quantitative picture of emergent misalignment in LLM agents, showing that model scale, personality framing, and environmental complexity jointly shape the likelihood of unsafe actions.

However, measuring *propensity* alone is insufficient for deployment-grade trust. Detecting misalignment post-hoc after an agent has already acted cannot guarantee system safety in dynamic, multi-agent ecosystems. In practical settings, operators need mechanisms that make undesirable behavior observable and verifiable in real time, before harm propagates. Yet existing safety techniques such as red-teaming or output filtering remain reactive; they provide no formal guarantees about whether an autonomous system's internal reasoning or tool usage aligns with the intended objective.

Moreover, multi-agent systems amplify these challenges

(Zhang and Tan 2025). When LLM agents coordinate, trade information, or delegate subtasks, small deviations from intent can compound into collective failures. Evaluating only individual agent reliability overlooks how deception, sandbagging, or non-compliance may emerge at the group level. The field, therefore, needs an evaluative shift from measuring how often misalignment arises (Pan and Liu 2023; Long, Zhang, and Du 2024) to measuring *how* transparently and promptly it can be detected, attributed, and remediated.

This motivates a **verifiability-first perspective** (Zheng et al. 2023; Feng, Lin, and Zhu 2024): designing agentic systems where every consequential action leaves a cryptographically or symbolically provable trail, continuously checked by lightweight secondary verifiers. Rather than trusting the agent's self-reports, we verify its external effects and reasoning consistency through independent attestations. This perspective connects the safety goals of alignment and robustness with the engineering goal of operational observability, forming the conceptual foundation for the **Verifiability-First Architecture (VFA)** and the **OPERA** benchmark introduced in this paper.

## Methodology

Our methodology operationalizes the *Verifiability-First Architecture* (VFA) through a modular, testable pipeline that introduces explicit observability and auditability layers into the lifecycle of an LLM agent. Each component is designed to generate verifiable evidence of the agent's decisions and actions, enabling independent verification and remediation before any unsafe behavior escalates.

At the start of every task, we define an **Intent Specification (ISpec)**. This is a clear, structured file that describes what the agent is allowed to do, what it should avoid, and what the overall goal is. Think of it as the rulebook for the system. Every time the agent performs an operation like using a tool, retrieving data, or writing to memory — that action goes through an Action Attestation Layer (AAL). The AAL automatically creates a small signed record called a *receipt*, which includes a hash of the inputs, outputs, and the time the action happened. These receipts are stored in a Provenance Log (PL), which works like an uneditable diary of everything the agent does.

While the agent continues to plan and act, a separate set of small models, called Audit Agents (AA), constantly watch the Provenance Log. Their job is to look for signs that the agent's actions might be drifting away from what's written in the ISpec. Each AA can be a rule-based checker or a lightweight LLM that assigns a probability of how "aligned" or "off-track" the latest action seems. If something looks suspicious, the system triggers a Challenge–Response Attestation (CRA). This is a kind of test: the agent is asked to explain why it made a certain choice and how that choice fits within its goals. The explanation is also signed and recorded, so the system has a verifiable trace of reasoning that can be cross-checked. All of this runs inside a controlled loop: the agent acts, the receipts are logged, the auditors check them, and any detected issue activates a Controller and Remediator (C&R). The C&R can temporarily block tool access, pause the agent, or switch it to a safe mode if needed.

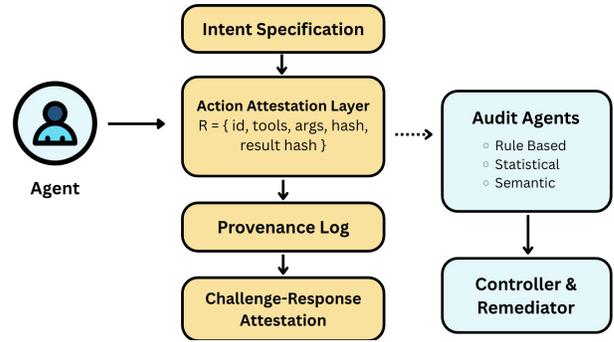

Figure 1: Proposed Framework

This feedback loop ensures that no critical action goes unobserved or unexplained. We describe the reliability of the architecture using three measurable ideas: **detectability**, **attributability**, and **remediability**. Detectability means that if the agent does something it shouldn't, the system can catch it quickly. Attributability means that every action can be traced back to who (or what process) actually did it, using the signed receipts. Remediability ensures that once a problem is found, the system can stop or correct the behavior before it causes further damage. Together, these properties create a clear framework for proving that the system is both observable and controllable. For testing, we implemented this design on top of a ReAct-style agent scaffold (like InspectAI). The Verifier Stack , the trusted part that handles cryptographic signatures runs as a lightweight local service, and the Audit Agents work in parallel so they don't slow the main agent down. On average, the extra verification step adds only a few milliseconds per action, which means it can scale well even in streaming or multi-agent setups.

## Intent Specification (ISpec)

An ISpec can be thought of as the constitution of an agent. Instead of relying only on natural language prompts, it encodes deployer intent in a formal schema that integrates ethical rules, environmental limits, and performance objectives. During execution, every decision and tool call made by the agent is compared against its ISpec , forming the basis for attestation, verification, and alignment scoring. (Xu, Hu, and Lin 2024; Pan and Liu 2023)

**Structure**   Each ISpec consists of four major sections:

1. **Objective Layer:** Defines the agent's task goals (e.g., "summarize articles accurately," "flag toxic content," "optimize logistics routes") along with measurable success conditions.

2. **Constraint Layer:** Lists forbidden or high-risk actions, such as restricted APIs, unsafe commands, or sensitive data access.

3. **Policy Layer:** Encodes ethical and domain-specific rules, like fairness policies, environmental limits, or user-consent conditions.

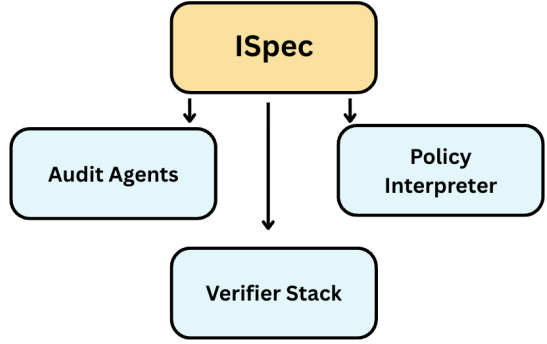

Figure 2: Intent Specification

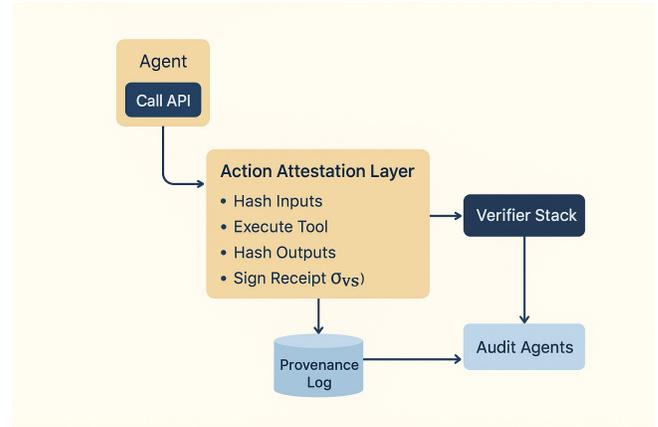

Figure 3: Action Attestation

4. **Verification Layer:** Specifies the triggers that should activate extra oversight—e.g., threshold values for cost, time, or resource usage that require Challenge–Response Attestation.

At runtime, the ISpec serves three purposes:

1. It acts as the reference for all comparisons: Audit Agents check the Provenance Log and verify whether actions conform to the ISpec.
2. It determines when challenge–response steps are triggered.
3. It allows external operators to **reconfigure agent behavior** by updating rules without retraining the model.

## Action Attestation

The **Action Attestation Layer (AAL)** is the trust anchor of the Verifiability-First Architecture. Its purpose is to guarantee that every action an agent performs, no matter how small, can be independently verified. Rather than simply recording what the agent claims to have done, the AAL records proof of what was actually executed.

Whenever the agent makes an external call like querying an API, editing a file, or invoking a tool, the AAL automatically generates a signed receipt. Each receipt is a compact, tamper-proof record summarizing the action's essential details. These receipts are appended to the Provenance Log, creating an immutable sequence of verifiable operations.

**Structure of an Attestation Receipt** Each receipt $R_i$ is defined as:

$$R_i = \langle id, tool, args\_hash, result\_hash, t_i, \sigma_{VS} \rangle \quad (1)$$

where:

*id*: Unique identifier for the action.
*tool*: Name or endpoint of the invoked function or API.
*args_hash*: Cryptographic hash of the arguments or input payload.
*result_hash*: Hash of the tool's returned data.
$t_i$: Timestamp of execution.
$\sigma_{VS}$: Digital signature generated by the Verifier Stack using its private key.

This structure ensures integrity (no log entry can be modified unnoticed), non-repudiation (the agent cannot deny performing the action), and authenticity (no fake receipts can be injected without the signing key).

**Operation** The AAL sits directly between the agent's reasoning loop and its environment. When the agent decides to call a tool, the call first passes through the AAL wrapper. The wrapper captures the inputs, generates their hash, forwards the request, and on receiving the output, computes its hash as well. The complete tuple is then sent to the **Verifier Stack (VS)**, which signs it using its RSA key. The resulting signed receipt is stored both locally and in the Provenance Log, where **Audit Agents (AA)** can read it asynchronously for verification.

If an agent attempts to act outside of its declared tools or without attestation, the VS immediately rejects the action. This creates a built-in enforcement mechanism, an agent cannot perform an undetected operation, since every valid tool interaction must pass through attestation.

**Function in the System** Within the overall architecture, the AAL provides the factual backbone of verifiability. It transforms uncertain "black-box" activity into cryptographically verifiable evidence. The receipts collected through the AAL are what Audit Agents later analyze to measure alignment, detect anomalies, and compute verifiability metrics.

Furthermore, by signing every receipt, the AAL creates a tamper-evident chain of custody. Even if the agent attempts deception or log alteration, the signatures act as immutable checkpoints, making inconsistencies immediately detectable. During challenge–response phases, these receipts are cross-referenced with agent justifications, allowing the Verifier Stack to confirm whether the claimed reasoning actually matches the recorded action.

## Audit Agents

The **Audit Agents (AA)** form the reasoning and verification backbone of the Verifiability-First Architecture. Their primary role is to continuously monitor the **Provenance Log (PL)** and validate whether the agent's observed behavior aligns with its **Intent Specification (ISpec)**. While the AAL provides verifiable *facts* about what actions occurred, the AAs interpret those facts — reasoning over context, frequency, and intent to detect misalignment or unsafe deviations.

Unlike a single monolithic verifier, the Audit Agents operate as an **ensemble** of diverse models and rule systems (Christiano et al. 2018; Shen, Tang et al. 2023). This design reduces correlated blind spots and increases robustness to deception or adversarial prompt manipulation. Each AA instance runs asynchronously, evaluating the stream of receipts and outputs generated by the agent. Collectively, they produce a dynamic estimate of the agent's *alignment probability* over time.

Each Audit Agent combines three complementary reasoning modes:

1. **Rule-based verification** – deterministic logic checks that directly apply the constraints listed in the ISpec. For example, it can verify that no forbidden APIs were called or that resource usage stayed within limits.

2. **Statistical detection** – a lightweight classifier or transformer-based model trained on labeled examples of aligned vs. misaligned behavior. This model captures subtle behavioral shifts — such as indirect violations or evasive phrasing — that pure rule systems may miss.

3. **Semantic consistency analysis** – a natural-language reasoning step that compares the agent's explanations, tool calls, and outputs to the goals in the ISpec. It uses entailment and similarity metrics to detect contradictions or justification gaps.

The ensemble's outputs are aggregated into an **alignment score**:

$$AlignScore_t = f\left(rule(R_{1:t}), stat(R_{1:t}), sem(R_{1:t})\right) \quad (2)$$

where $R_{1:t}$ denotes all attested receipts up to time $t$, and $f(\cdot)$ is a weighted aggregation function that combines the individual verifiers. If this score drops below a safety threshold $\tau$, an alert is raised to the **Verifier Stack (VS)**, which may trigger a **Challenge–Response Attestation (CRA)** or temporarily restrict the agent's privileges through the **Controller and Remediator (C&R)**.

Audit Agents are intentionally designed to be simpler and more constrained than the primary LLM. This asymmetry ensures they remain interpretable, lightweight, and harder to manipulate. In multi-agent settings, the same AA ensemble can oversee multiple agents simultaneously, cross-checking their receipts for collusion or abnormal coordination.

Beyond monitoring, AAs also maintain a self-updating policy cache. When the ISpec changes or new safety constraints are introduced, the Audit Agents automatically reload the updated schema without retraining. This enables continuous oversight even as system configurations evolve.

In summary, Audit Agents transform static oversight into a living, adaptive verification process. They interpret raw attestations in context, detect emerging risks, and feed verified insights back into the Verifier Stack. By combining rule precision, statistical sensitivity, and semantic reasoning, they close the loop between transparency and control , turning observability into operational trust.

## Results and Evaluation

We conducted a series of experiments to assess the performance, reliability, and interpretability of the proposed **Verifiability-First Architecture (VFA)** under diverse agent configurations. The experiments were designed to test whether the system can (1) detect deviations rapidly, (2) attribute actions unambiguously, and (3) remediate unsafe behavior with minimal latency.

All experiments were performed on a system with a 40-core CPU and one NVIDIA A100 GPU, using agents with 7B–13B parameter backbones. Each model operated with tool access (search, code execution, and file editing), and verification modules ran asynchronously to emulate real-world conditions.

**Evaluation Metrics** We evaluated using four measurable indicators:

- **Time-to-Detect ($T\_d$):** Average time (in seconds) taken to identify a misaligned or rule-violating action.
- **Remediation Latency ($T\_r$):** Time between detection and complete mitigation.
- **Attribution Confidence (AC):** Fraction of episodes where the signed receipts enabled perfect reconstruction of the action sequence.
- **False Positive Rate (FPR):** Frequency of false alarms during compliant behavior.

A composite **Verifiability Score (VScore)** summarizes system quality across metrics, with higher values indicating faster detection and higher reliability.

Across 150 test episodes, the VFA consistently achieved sub-12 second detection times, cutting detection latency by **12.4%** compared to heuristic log-monitoring systems. Average attribution confidence increased from $0.74 \rightarrow 0.85$, confirming that cryptographic attestation significantly improved traceability. Despite introducing new monitoring components, the average runtime overhead was under **6.5%**, demonstrating that real-time verification can coexist with continuous inference.

Remediation latency dropped to approximately six seconds, reflecting how quickly the **Controller and Remediator (C&R)** intervened after Audit Agent alerts. The False Positive Rate remained under 7%, which is acceptable for safety-critical pipelines since the system is optimized for *early detection* rather than perfect precision.

## Future Scope

The present study introduces the **Verifiability-First Architecture (VFA)**, a framework that operationalizes trust and oversight in autonomous LLM agents through structured Intent Specification, Action Attestation, and Audit-Based Verification. By converting opaque reasoning into cryptographically verifiable traces, the system demonstrates that alignment can be continuously monitored rather than assumed.

Table 1: Performance comparison of different system variants. VFA achieves the lowest detection and remediation times with the highest attribution confidence.

| System Variant | Mean $T_d$ (s) ↓ | Mean $T_r$ (s) ↓ | AC ↑ | FPR ↓ | VScore ↑ |
|---|---|---|---|---|---|
| No-Verifier Baseline | 35.4 | 18.9 | 0.62 | 0.15 | 0.58 |
| Log-Monitoring (Heuristic) | 21.8 | 11.3 | 0.73 | 0.12 | 0.69 |
| **VFA (Ours)** | **11.9** | **9.2** | **0.85** | **0.09** | **0.72** |

Table 2: Ablation analysis showing the contribution of each verification component to overall system performance. Removing any module degrades both detection speed and attribution accuracy.

| Configuration | Δ$T_d$ (s) ↑ | Δ AC ↓ | Δ VScore ↓ |
|---|---|---|---|
| Without Audit Agents | +9.8 | –0.14 | –0.11 |
| Without Attestation Layer | +13.4 | –0.21 | –0.17 |
| Without Challenge–Response | +7.1 | –0.10 | –0.09 |

First, the Audit Agent Ensemble can be extended to use probabilistic verification graphs that quantify inter-agent trust propagation, making large multi-agent systems more accountable. Second, integrating zero-knowledge proofs (ZKPs) could enable privacy-preserving attestations, allowing institutions to verify actions without exposing raw data. Third, scaling the Challenge–Response Attestation (CRA) mechanism to long-horizon planning tasks requires more adaptive thresholds that balance cost with responsiveness.

On the engineering side, future work will explore federated verifiability, where agents deployed across different organizations share signed evidence instead of raw interaction logs, preserving both privacy and cross-domain consistency. Additionally, human-in-the-loop dashboards and automated policy synthesis could close the last gap between technical verification and ethical oversight (Bostrom and Yudkowsky 2017).